\documentclass[12pt]{iopart}

\usepackage{graphicx,epsfig}

\newcommand{\eins}{1\!\! 1}                  
\renewcommand{\e}{\mathrm{e}}                
\renewcommand{\i}{\mathrm{i}}                

\begin{document}

\title{The intensity correlation function of ``blinking'' quantum systems}
\author{Gerhard C. Hegerfeldt and Dirk Seidel}
\address{Institut f\"ur Theoretische Physik, 
Universit\"at G\"ottingen, Germany}
\ead{hegerf@theorie.physik.uni-goettingen.de}

\begin{abstract}
Explicit expressions are determined for the photon correlation
function of ``blinking'' quantum systems, i.e.~systems with different types of
fluorescent periods. These expressions can be
used for a fit to experimental data and for obtaining system
parameters therefrom. For two dipole-dipole interacting $V$ systems the
dependence on the dipole coupling constant is explicitly given and
shown to be particularly pronounced if the strong driving is
reduced. We propose to use this for an experimental verification of
the dipole-dipole interaction. 
\end{abstract}
\submitto{\JOB}

\section{Introduction}

For fluorescing quantum systems one of the most important statistical
quantities is given by the intensity correlation function, $g(\tau)$,
for photon counts 
\cite{Mandel-book}. Its behavior for small times $\tau$ can indicate
more classical 
or more quantum behavior and bunching or anti-bunching, depending on whether
$g(0)>1$ or $g(0)<1$. Early  investigations of the intensity
correlation function of single two level systems
\cite{Carmichael-JP-1976, Kimble-PRA-1976,Agarwal-PRA-1977} led to the 
observation of nonclassical light
\cite{Kimble-PRL-1977,Kimble-PRA-1978}. The intensity correlation 
function can also exhibit Rabi oscillations, and for $V$ systems with
metastable state it  contains 
indications of light and dark periods (''blinking'') in the
fluorescence  \cite{Plenio-RMP-1998}.

The importance of correlation functions partially stems from the relative
ease with which they can be experimentally determined, partially due
to the fact that the efficiency of the photon detector does not
enter. Correlation functions were determined for single ions in a
trap in  experiments which were mainly carried out for the detection
of quantum jumps in the fluorescence
\cite{Sauter-OC-1986, Berquist-PRL-1986,Diedrich-PRL-1987,Itano-PRA-1988}.
In the last decade the same was achieved for single
fluorescent molecules 
\cite{Basche-Nat-1995,Basche-JPC-1995,Basche-BBG-1996}. A theoretical
determination of $g(\tau)$ was obtained in closed form
for three level systems \cite{Agarwal-ZP-1979,Pegg-PRA-1986,
  Nienhuis-PRA-1987} and $g(0)$ was calculated for two dipole-dipole
interacting two-level systems \cite{Richter-OA-1982, Ficek-PRA-1984,
  Beige-PRA-1998}, but for more 
complicated multi-level systems it is usually done by the numerical
solution  of Bloch equations \cite{Agarwal-PRA-1977,
  Lawande-PRA-1989,Lawande-PRA-1990,Ficek-PRA-1990, Schubert-PRA-1995}. Because the
intensity correlation function contains information on the system parameters,
e.g.~atomic constants like Einstein coefficients, these
parameters could be obtained in principle from an experimentally
determined $g(\tau)$ by fitting numerically calculated curves to the
data. However, in general such a fit procedure needs many numerical
runs and tends to be highly sensitive to experimental and numerical
errors. With an  algebraic expression a fit would be much
easier and more 
reliable. Moreover, if algebraic expression were known for the 
intensity correlation function for multi-level systems one could study its
behavior without recourse to many numerical runs for different
parameter values. In the present
paper such algebraic expressions of the correlation function will be given for
fluorescing quantum systems with light and dark periods. 
As a generalization of a  result for $V$ systems in
\cite{Plenio-RMP-1998}  it is deduced
from this algebraic expression that the correlation function for
``blinking'' systems with dark periods shows a ``hump''  
for values of $\tau$ larger than the correlation times of the
individual intensity periods.  
It is also   pointed out that the existence of extended dark
periods may considerably enhance the amplitude of Bloch oscillations. 

As an application  we study two
dipole-dipole interacting $V$ systems and determine  an algebraic
expression for the corresponding intensity correlation function
 $g(\tau)$. This expression is used to  study the
behavior of $g(\tau)$ with respect to the strength of the dipole-dipole
interaction. It is 
shown that the dipole-dipole interaction has an effect   for small
$\tau$ which is particularly pronounced for small values of 
the strong driving. We suggest to use this effect  to experimentally
verify the dipole-dipole interaction for atomic
distances of a few wavelength of the strong transitions.

\section{Algebraic expressions for intensity correlation functions}

The temporal intensity correlation function for photon counts is defined as
follows. In the steady state, let $G(\tau)$ be the joint probability
density for detecting a photon both at time 0 and $\tau$, and let
$I_{\rm SS}$ denote the steady-state intensity (counts per unit time). Then
$g(\tau)$ is defined as \cite{Carmichael-book}
\begin{equation}
g(\tau)= G(\tau)/I^2_{\rm SS}~.
\end{equation}
We consider now a fluorescing system with $n$ periods of different intensity
$I_i$ and mean duration $T_i, i=0,\dots,n-1$. A particular period usually
corresponds to transitions within a simpler subsystem which is
easier to treat than the complete system. In the following we
therefore assume that within a given period $i$ the corresponding
intensity correlation function $g_i(\tau)$ is known. 

The intensity correlation function $g(\tau)$ of the larger system can then be
determined as follows.
Let  $P_i$ be the probability for the occurrence of period $i$
 and let $P_{ij}(\tau)$ be the probability to have period
$j$ at time $\tau$ provided one had period $i$ at $\tau = 0$. Now, if at
time 0 the fluorescence is in period $i$, then, for $\tau \ll T_i$,
one still is in period $i$ with high probability. Therefore, in this
case, the joint probability density for detecting a photon at both
times, 0 and $\tau$, is $\sum_i P_i G_i (\tau)$ where $G_i (\tau)$ is
the joint  probability density for period $i$. Since for small $\tau$ one has
$P_{ii}(\tau) \simeq 1$ this can be replaced by
\begin{equation} \label{eq:small_tau}
  \sum_i P_i P_{ii}(\tau) G_i(\tau)~,\qquad \tau\ll T_i~.
\end{equation}
For larger $\tau$ one may end up in a period different from that at
time 0. If $\tau$ is larger than the correlation times associated with
$g_i(\tau)$ then,  for initial period $i$
and final period $j$, the joint
detection probability is $I_i P_{ij}(\tau) I_j$. Since for such
$\tau$'s one has $ G_j(\tau)=I_j^2$  one can write the complete joint
detection probability density as
\begin{equation} \label{eq:big_tau}
  G(\tau) = \sum_{ij} P_i I_i P_{ij}(\tau) G_j(\tau)/I_j~.
\end{equation}
This includes equation \eref{eq:small_tau} as a special case since for
$\tau$ small one has $P_{ij} (\tau) \cong 0$ for $i \ne j$.

In this general formula one still has to express $P_i$ and
$P_{ij}(\tau)$ in terms of the transition rates $p_{ij}$ from period $i$ to
period $j$. These transition rates themselves can be expressed by Einstein
coefficients and Rabi frequencies. One has, in particular,  $T_i = 1/
\sum_{k \ne i} p_{ik}$. The $P_{ij}(\tau)$ are easily seen
to obey rate equations, e.g.
\begin{equation}
  \dot{P}_{11}(\tau) = \Bigl(-\sum_k p_{1k}\Bigr)P_{11}(\tau) +
    p_{21}P_{12}(\tau)+ \cdots + p_{n1}P_{1n}(\tau)~.
\end{equation}
In general, with the matrix ${\bf B} = (B_{ij})$,
\begin{equation}
 B_{ij}=p_{ij} -
 \delta_{ij}\sum_{k}p_{ik}~,  
\end{equation}
and the matrix
\begin{equation}
  {\bf P}(\tau) = \Bigl(P_{ij}(\tau)\Bigr)
\end{equation}
one has
\begin{equation} \label{eq:P_diffeq}
 \dot{\bf P} = {\bf P}{\bf B}~, 
\end{equation}
with the initial condition $P_{ij}(0) = \delta_{ij}$, or
\begin{equation}
 {\bf P}(0) = \eins~. 
\end{equation}
The solution of equation \eref{eq:P_diffeq} with this initial condition
can be written as
\begin{equation} \label{eq:P_tau}
 {\bf P}(\tau) = \e^{{\bf B}\tau}~. 
\end{equation}
If $\mu_0, \dots, \mu_{n-1}$ are the eigenvalues of ${\bf B}$ (assumed
distinct) then \cite{gant}
\begin{equation} \label{eq:gantm_formel}
  \e^{{\bf B}\tau} = \sum_{i=0}^{n-1}\e^{\mu_i\tau}\prod_{\alpha\neq
    i}\frac{{\bf B}-\mu_{\alpha}}{\mu_i-\mu_{\alpha}}~.
\end{equation}
The properties of the matrix ${\bf B}$ are closely related to those of
stochastic matrices \cite{gant,stoch_matrix}, and under quite
general conditions ${\bf B}$ has a single eigenvalue $\mu_0 = 0$ and
eigenvalues $\mu_1,\dots, \mu_{n-1}$ with negative real part. This
will be assumed in the following. For up to four intensity periods,
the $\mu_i$ can be determined in closed form. 

To find the $P_i$'s, we note that for $\tau \rightarrow \infty$ the
memory to the initial start is in general lost. Therefore, for any $\kappa$,
\begin{equation} \label{eq:P_i}
  P_i = P_{\kappa i}(\infty)~.
\end{equation}
Laplace transforming equation \eref{eq:P_tau} yields
\begin{equation}
  {\bf P}(\infty) = \lim_{\epsilon\rightarrow +0}\,\epsilon(\epsilon -
  {\bf B})^{-1}~.
\end{equation}
Thus the $P_i$'s can be calculated without knowledge of the $\mu_i$'s.

The steady-state fluorescence intensity  is $I_{\mathrm{SS}} = \sum_i
P_i I_i$~. Therefore we obtain
\begin{equation} \label{eq:gt}
  g(\tau) = \frac{G(\tau)}{I_{\mathrm{SS}}^2} = \frac{\sum_{ij} P_i
    I_i I_jP_{ij}(\tau) g_j(\tau)}{\left(\sum_{\alpha}
      P_{\alpha}I_{\alpha}\right)^2}~. 
\end{equation}
For $\tau$ larger than the correlation time of $g_i (\tau)$ one has
$g_i(\tau)\simeq 1$, and thus equation \eref{eq:gt} becomes
\begin{equation} \label{eq:gt_modell}
  g(\tau) = \frac{\sum_{ij}P_i I_iI_j P_{ij}(\tau)}{\left(\sum_{\alpha}
    P_{\alpha} I_{\alpha}\right)^2}
\end{equation}
so that for larger $\tau$ the dependence on  $\tau$ of $g(\tau)$ is solely
governed by the statistics of the individual periods.

\begin{figure}[htbp]
  \begin{center}
    \epsfxsize=7cm  \epsfbox{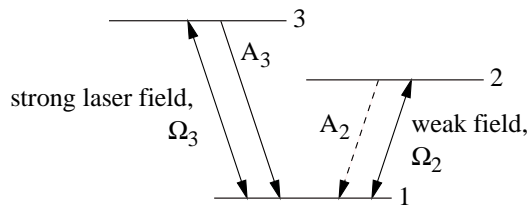}
    \caption{$V$ system with metastable level 2 and Einstein coefficient
      $A_3$ for level 3. $\Omega_2$ and $\Omega_3$ are the Rabi
      frequencies of the two lasers driving the weak 1-2 transition
      and the strong 1-3 transition, respectively.}
    \label{fig:V-schema}
  \end{center}
\end{figure}

A first simple example is a system with  dark ($I_0 = 0$)
and  light ($I_1\ne 0$) periods, such as  the $V$ system of \fref{fig:V-schema}
 with a metastable level   or a $\Lambda$ system 
\cite{Itano-PRA-1988, Dehmelt-BAP-1975,Cook-PRL-1985, Cohen-EPL-1986,
  Schenzle-PRA-1986, Itano-PRL-1987}. The  mean period durations are  $T_0 =
1/p_{01}$ and $T_1 = 1/p_{10}$, respectively.
Because $I_0 = 0$, equation \eref{eq:gt} becomes
\begin{equation} \label{eq:gt_3NS}
  g(\tau) = \frac{1}{P_1} P_{11}(\tau) g_1(\tau)~,
\end{equation}
with $g_1(\tau)$  the correlation function of the two-level
subsystem. A simple calculation yields
\begin{equation}
  P_{11}(\tau) = \frac{T_1}{T_0 +T_1} + \frac{T_0}{T_0 +
    T_1}\e^{-\bigl(\frac{1}{T_0} + \frac{1}{T_1}\bigr)\tau}
\end{equation}
\begin{equation}
  P_1 = \frac{T_1}{T_0 + T_1}~.
\end{equation}
For a $V$ system equation \eref{eq:gt_3NS} agrees with the result 
of  \cite{Nienhuis-PRA-1987} if the correlation time for $g_1(\tau)$ is
much smaller than $T_1$.

For $\tau\ll T_1$ one has $P_{11}(\tau)=
1$  and  $g(\tau)= g_1 (\tau)/P_1$. Since $P_1
< 1$  it follows that, for small $\tau$, $g(\tau)$ is just $g_1(\tau)$ 
blown up by the factor $1/P_1$. In particular, possible Bloch
oscillations of $g_1(\tau)$ become enhanced in 
$g(\tau)$ if $T_0$ increases. Moreover, for $\tau$ values
larger than the correlation time of the two-level subsystem one has
$g_1 (\tau)=1$ so that there is a
``hump'' larger than 1 in $g(\tau)$. In the case of a $V$ system this
hump was already noted in \cite{Plenio-RMP-1998}.  

Such a hump is a {\em general feature} for any system with a dark
period, as will now be shown by means of
equation \eref{eq:gt_modell}.  Indeed, for $\tau$ larger than the
correlation times of $g_i(\tau)$ but much less than the $T_i$'s one
has $P_{ij}= \delta_{ij}$, resulting in  $g(\tau)=\sum_i P_i
I_i^2/(\sum_{\alpha} P_{\alpha} I_{\alpha})^2$. By Schwarz's
inequality one obtains $(\sum_{\alpha} P_{\alpha}^{1/2}
P_{\alpha}^{1/2} I_{\alpha})^2\leq
(\sum_{\alpha \neq 0}P_{\alpha})(\sum_{\alpha}
P_{\alpha}I_{\alpha}^2)$. Since $\sum_{\alpha \neq 0}P_{\alpha}= 1 -
P_0 < 1$ the statement follows. This does not mean, however, that
$g(\tau)$ always stays above 1 for all subsequent values of
$\tau$. There are systems where it dips below 1 again and then
approaches its asymptotic value 1 from below.

In the following, equation \eref{eq:gt} for $g(\tau)$ will be
studied for an example with three different fluorescence periods and
shown to be highly accurate. 

\section{Application to two dipole-dipole interacting atoms} \label{kap:2DNS}

We consider two dipole-dipole interacting $V$ systems as in
\fref{fig:V-schema} at a
fixed distance $r$, with one laser driving the strong 1-3 transition
and another the weak 1-2 transition. Such a system exhibits
three fluorescence periods, a dark period $I_0 = 0$, a period $I_1$
and a double intensity period $I_2 \simeq 2 I_1$. The
transition rates $p_{ij}$ between the periods are known
\cite{Addicks-EPJ-2001} and are given in the Appendix. One has $T_0 =
1/p_{01}$, $T_1 = 1/(p_{10} + p_{12})$,  $T_2 = 1/p_{21}$, and $p_{02}=
p_{20} = 0$. The  
distance dependent complex dipole-dipole coupling constant $C_3$ is also given
in the Appendix. 
For $r\rightarrow\infty$ the system behaves as two independent,
non-interacting, fluorescing $V$ systems, and then the periods $I_2$,
$I_1$ and $I_0$ correspond to two, one or no atom radiating.
Because of $I_0=0$, equation \eref{eq:gt} simplifies to
\begin{eqnarray} \label{eq:gt_2DNS}
  \fl g(\tau) = \Bigl\{P_1 I_1^2 P_{11}(\tau) g_1(\tau) + P_1 I_1 I_2
    P_{12}(\tau) g_2(\tau) + P_2 I_1 I_2 P_{21}(\tau) g_1(\tau)
     + P_2 I_2^2 P_{22}(\tau)
    g_2(\tau)\Bigr\} \Big/ \nonumber\\ (P_1 I_1 + P_2
    I_2)^2~.
\end{eqnarray}
Here, $g_1 (\tau)$ is the usual correlation function of a single
two-level system and $g_2(\tau)$ that of two two-level systems which
are dipole-dipole interacting. These correlation functions are given
in  the Appendix. 
The transition probabilities $P_{ij} (\tau)$ from period $i$ to period
$j$ in time $\tau$ are easily calculated from equation \eref{eq:P_tau}
and are also 
given in the Appendix, as is $P_i$, the probability for the occurrence
of period $i$. 

\begin{figure}[htbp]
  \begin{center}
    \epsfxsize=11cm  \epsfbox{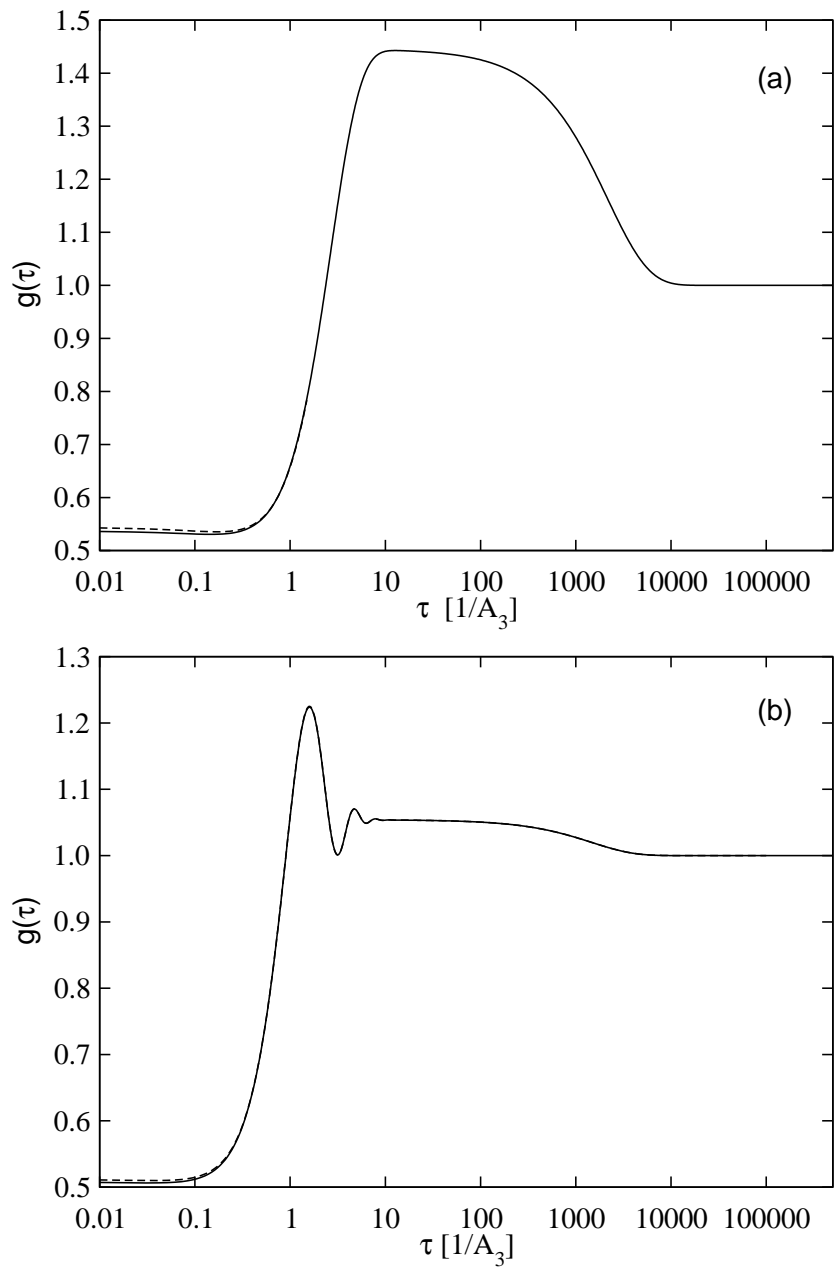}
    \caption{$g(\tau)$ for two dipole-interacting $V$ systems with an
      atomic distance $r=2.7\,\lambda$. The result of equation
      \eref{eq:gt_2DNS} 
      (solid line) is compared with a numerical calculation (dashed
      line). The hump after $\tau > 10 \,A_3$ is clearly visible. The
      difference for small $\tau$ comes from the the 
      restriction to first-order terms in $C_3$. (a)
      $\Omega_3=0.3\,A_3$ and  $\Omega_2=0.005\,A_3$; 
      (b) $\Omega_3=5.0\,A_3$ and $\Omega_2=0.05\,A_3$.}
    \label{fig:gt_9niv_0.3_0.005}
  \end{center}
\end{figure}

The accuracy of equation \eref{eq:gt_2DNS} is checked for two values of
$\Omega_3$ in
\fref{fig:gt_9niv_0.3_0.005} for a coupling constant $\mathrm{Re}\,C_3 = -0.09\,A_3$. If the strong
laser and the atomic dipole moments are perpendicular to the atomic
connecting line this corresponds 
to a local maximum at an atomic distance $r = 2.7\,\lambda$, where
$\lambda$ is the wave length of the strong 3-1 transition. The agreement is
excellent and becomes even better if one takes higher orders of
$C_3$ into account, instead of only the first order as done here. 

The usefulness of equation \eref{eq:gt_2DNS} is twofold. First, it can be
used for a 
fit to experimental data to obtain atomic parameters and mean period
lengths. Experimentally, correlations functions are in general much
easier to determine than period lengths.
For such a  fit one could also employ numerical solutions for $g
(\tau)$ obtained with the quantum regression theorem \cite{Lax} or with the
quantum jump approach
\cite{Hegerfeldt-proceedings,Hegerfeldt-PRA-1993,Hegerfeldt-QSO-1996,
  Beige-PRA-1998}. However, the numerical
approach is in general much 
more sensitive to experimental errors than a fit based on an analytic
expression.

The second use of equation \eref{eq:gt_2DNS} lies in the feasibility
to study the  behavior of $g(\tau)$ for all
parameter values simultaneously, without having to perform many numerical
runs and possibly overlooking
interesting parameter values. This will be demonstrated here by exhibiting a
possible experimental test of the dipole-dipole interaction.

\section{Possible experimental verification of the dipole-dipole interaction}
\label{kap:exp}

\begin{figure}[htbp]
  \begin{center}
    \epsfxsize=11cm  \epsfbox{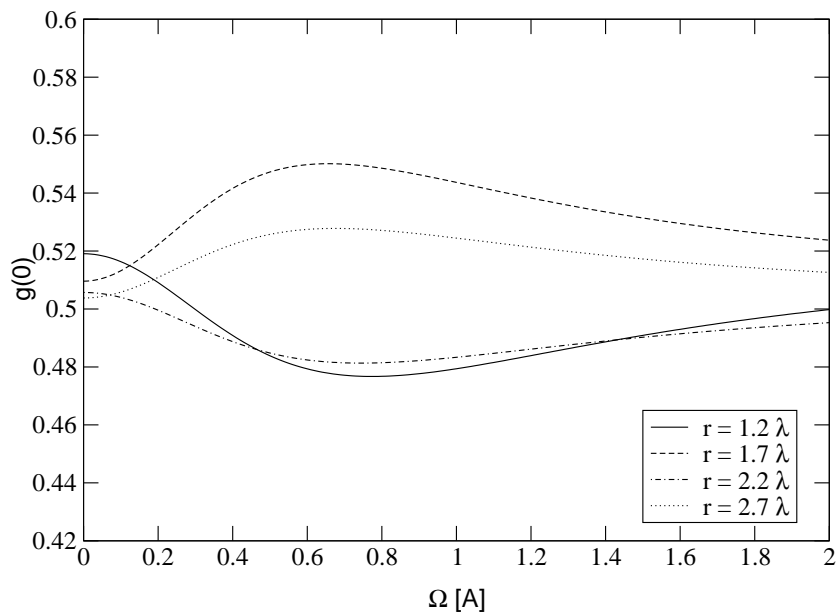}
    \caption{$g(0)$ of equation \eref{eq:g2(0)} for two dipole-interacting
      two-level systems.} 
    \label{fig:g0_2ZNS}
  \end{center}
\end{figure}

For $\tau\ll T_1,T_2$, one has $P_{ij}(\tau)= \delta_{ij}$, so that
equation \eref{eq:gt_2DNS} reduces to
\begin{equation} \label{eq:gt_2DNS_smalltau}
  g(\tau) = \frac{P_1 I_1^2 g_1(\tau) + P_2
    I_2^2 g_2(\tau)}{(P_1 I_1 + P_2 I_2)^2}~.
\end{equation}
In particular, one has $g_1(0) = 0$, and $g_2(0)$ for angle-averaged
detection \cite{Skornia} is given in
equation \eref{eq:g2(0)} of the Appendix. Inserting this and the other
quantities into equation \eref{eq:gt_2DNS_smalltau} one obtains
\begin{equation} \label{eq:g0_2DNS}
  g(0) = \frac{2P_2(A_3^2+2\Omega_3^2)(A_3^2+(\mathrm{Re}\,
    C_3)^2)N}{A_3^2(P_1 N +
    2 P_2 (A_3^2+2\Omega_3^2)(A_3^2 + 2\Omega_3^2 +
    A_3 \mathrm{Re}\,C_3))^2}
\end{equation}
with $N=(A_3^2+2\Omega_3^2)^2 + A_3^2 |C_3|^2 + 2A_3^3\mathrm{Re}\,C_3$.
To first order in the coupling constant $C_3$ this becomes
 \begin{equation}
  g(0) = \frac{1}{2} - \frac{A_3}{2}\frac{(A_3^2+\Omega_3^2)^2 +
    \Omega_3^4}{(A_3^2 + \Omega_3^2)^2(A_3^2 + 2\Omega_3^2)}\mathrm{Re}\,C_3~.
\end{equation}
By varying $\Omega_3$ one has a noticeable change of $g(0)$, which
depends on how large $\mathrm{Re}\,C_3$ is. For small $\tau$, the temporal
behavior of $g(\tau)$ is a combination of that of $g_1(\tau)$ and
$g_2(\tau)$ and can be experimentally resolved. Therefore $g(0)$ should
be measurable and the presence of the dipole-dipole interaction
detectable. The deviation of $g(0)$  from $1/2$, i.e. from that for
non-interacting atoms, is greatest for small $\Omega_3$ and drops off
to zero for increasing $\Omega_3$. 
\begin{figure}[htbp]
  \begin{center}
    \epsfxsize=11cm  \epsfbox{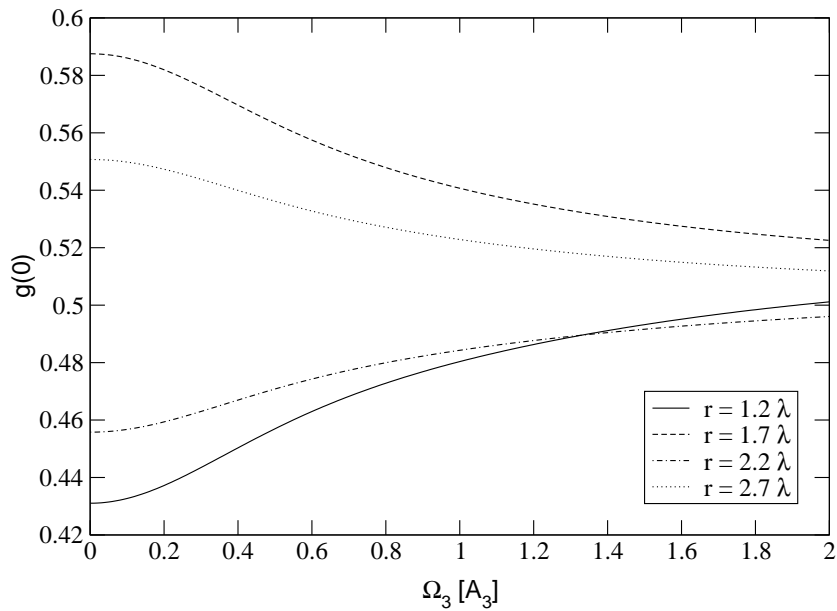}
    \caption{$g(0)$ of equation \eref{eq:g0_2DNS} for two
      dipole-interacting $V$ systems.}  
    \label{fig:g0_2DNS}
  \end{center}
\end{figure}
In \fref{fig:g0_2DNS} we
have plotted $g(0)$ for various values of $\Omega_3$ and for
$\mathrm{Re}\,C_3 = 
0.2\,A_3,~-0.1\,A_3,~0.1\,A_3,~-0.09\,A_3$. If the strong
laser and the atomic dipole moments are perpendicular to the atomic connecting line this corresponds to the
atomic distance $r$ 
=1.2$\,\lambda$, 1.7$\,\lambda$, 2.2$\,\lambda$, 2.7$\,\lambda$,
respectively. Since  
$\mathrm{Re}\,C_3$ oscillatingly drops off 
to zero for increasing atomic distance, its influence on $g(0)$
diminishes for increasing atomic distance. As shown in
\fref{fig:g0_2ZNS} the effect is about a factor of two smaller in
the case of two dipole-dipole interacting two-level systems.

\begin{figure}[htbp]
  \begin{center}
    \epsfxsize=11cm  \epsfbox{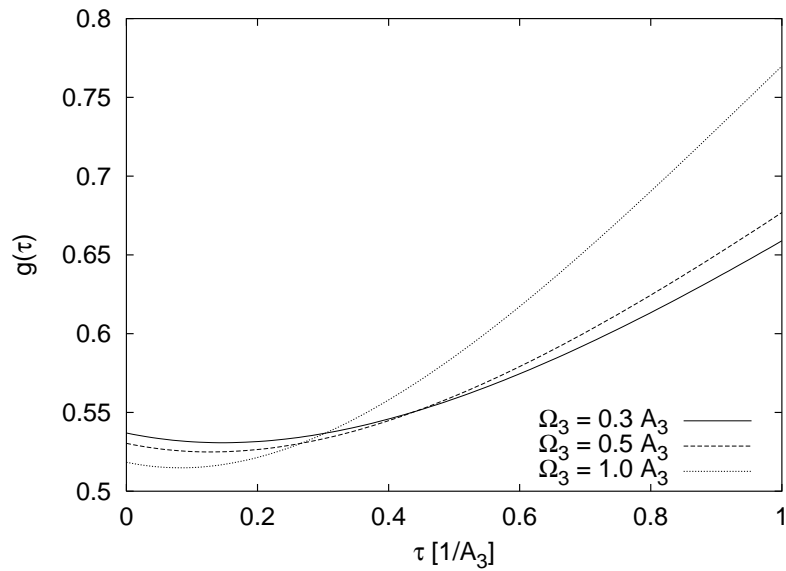}
    \caption{$g(\tau)$ to first order in $C_3$ for two
      dipole-interacting $V$ systems with an atomic distance of
      $r=2.7\,\lambda$ and $\Omega_2=0.005\,A_3$.}
    \label{fig:gtkl}
  \end{center}
\end{figure}

To show that the temporal behavior of $g(\tau)$ should  be
experimentally resolvable  we have plotted $g(\tau)$ in
\fref{fig:gtkl} for various
values of $\Omega_3$ and for $\mathrm{Re}\,C_3 = -0.09\,A_3$, which corresponds
to an atomic distance of $r=2.7\,\lambda$ for perpendicular laser incidence. 

\section{Discussion}

It has been shown for systems with different fluorescent periods that their
intensity correlation function can be reduced to those of simpler subsystems
and to quantities that govern the stochastic behavior of the different
periods. This is a considerable simplification and allows the determination of
an algebraic expression for the intensity correlation function which usually
have to be calculated numerically via the quantum regression theorem or via
the quantum jump approach.

One of the advantages of an algebraic expression is the feasibility of
studying its behavior for all parameters simultaneously, without having to
make numerical runs for different  sets of parameters and possibly overlooking
interesting values. Another advantage is that it is easier to fit experimental
data to an algebraic expression and to obtain unknown atomic
parameters or period durations by such a fit rather
than  fitting to numerically determined expression since the latter
procedure is usually much more sensitive to experimental and numerical errors.

It has been shown in this paper that the correlation function for
``blinking'' systems with dark periods shows a ``hump'' larger than 1 
for values of $\tau$ larger than the correlation times of the
individual intensity periods. This generalizes a similar result for a
$V$ system  in  \cite{Plenio-RMP-1998}. 
It has also  been pointed out that the existence of extended dark
periods may considerably enhance the amplitude of Bloch oscillations. 

As an application we have derived an expression for the intensity correlation
function of two dipole-dipole interacting $V$ systems and have studied its
behavior with respect to the strength of the dipole-dipole
interaction. We have 
shown that there is a significant trace of the dipole-dipole interaction in
$g(\tau)$ for small $\tau$ which is particularly pronounced for small values of
the strong driving. This effect of the dipole-dipole
interaction should be experimentally verifiable for atomic
distances of a few wavelength of the strong transitions. For two dipole-dipole
interacting two-level systems the effect is about a factor of two smaller.

\begin{appendix}
\section{}
The temporal intensity correlation function for a single two-level system is
given by \cite{Carmichael-JP-1976}
\begin{equation}
g_1 (\tau) = 1- \e^{\textstyle
  -\frac{3}{4}A\tau}\left(\cos\gamma\tau + \frac{3A}{4\gamma}\sin
  \gamma\tau\right) 
\end{equation}
with $\gamma = \sqrt{16\Omega^2-A^2}/4$, and the mean intensity is
\begin{equation}
I_1 = \frac{A \Omega^2}{A^2 + 2\Omega^2}~.
\end{equation}
We have used the approach of   \cite{Beige-PRA-1998} and Eq.~(27) of
  \cite{Hegerfeldt-PRA-1993}  to determine the corresponding expression for
two dipole-dipole interacting two-level systems and find
\begin{eqnarray}
\fl g_2(\tau) = 1 - \frac{1}{2}\e^{\textstyle
  -\frac{3}{4}A\tau}\left(\cos\gamma\tau + \frac{3A}{4\gamma}\sin
  \gamma\tau\right) \nonumber\\ 
 - \frac{1}{2}\frac{A\,\mathrm{Re}\,C}{(A^2+2\Omega^2)^2}\,\e^{\textstyle
  -\frac{3}{4}A\tau} 
\Biggl[\Biggl(4\Omega^2+\frac{A(A^2+2\Omega^2)(A^2 -
  22\Omega^2)}{16\gamma^2}\,\tau\Biggr) \cos \gamma\tau\nonumber\\
 - \Biggl(\frac{512\Omega^6+41A^6+2A^2\Omega^2(776\Omega^2-391A^2)}{
  64A\gamma^3} + \frac{(A^2-6\Omega^2)(A^2+2\Omega^2)}{4\gamma}\,
\tau\Biggr) \sin \gamma\tau\Biggr]\nonumber\\
\quad
\end{eqnarray}
to first order in the dipole-dipole coupling constant $C$. To all
orders in $C$ one has \cite{Ficek-PRA-1984,Beige-PRA-1998}
\begin{equation} \label{eq:g2(0)}
 g_2(0) = \frac{A^2+(\mathrm{Re}\,
   C)^2}{2A^2}\left[1+\frac{A(A\,(\mathrm{Im}\,C)^2-4\Omega^2\mathrm{Re}\, 
      C)}{(2\Omega^2+A(A+\mathrm{Re}\,C))^2}\right] 
\end{equation}
and
\begin{equation}
I_2 = \frac{2A\Omega^2(2\Omega^2 + A(A+\mathrm{Re}
    \,C))}{(A^2+2\Omega^2)^2+A^2\mathrm{Re}\,C(2A+\mathrm{Re}\,C) +
    A^2(\mathrm{Im}\,C)^2}~.
\end{equation}

The dipole-dipole coupling constant $C$ is given by
\cite{Beige-PRA-1998,Beige-PRA-1999,Agarwal-book} 
\begin{equation}
 C = \frac{3A}{2}\e^{\i kr}\Biggl[\frac{1}{\i
    kr}\left(1-\cos^2\vartheta\right)
  + \left(\frac{1}{(kr)^2}-\frac{1}{\i
      (kr)^3}\right)\left(1-3\cos^2\vartheta\right)\Biggr]~,
\end{equation}
where $\vartheta$ denotes the angle between the dipole moments of the
atoms and their connection line and $k$ is the wave number of the
strong transition. We assume $\vartheta=\pi/2$ for maximal values of $C$.
In Sections \ref{kap:2DNS} and \ref{kap:exp} these expressions are
used with $A = A_3, \Omega = \Omega_3$ and $C = C_3$.

For systems with two different light periods and a dark period the
relevant transition probabilities, $P_{ij}(\tau)$,  from period $i$ to
period $j$ in  
time $\tau$ are found from equation \eref{eq:P_tau}. For physical
reason it is assumed that $p_{02}$ and $p_{20}$ vanish. The
eigenvalues of the matrix ${\bf B}$ are then  $\mu_0 = 0$ and
\begin{equation}
\fl \mu_{1,2} = -\frac{1}{2}(p_{01}+p_{10}+p_{12}+p_{21})\pm
\frac{1}{2}\sqrt{(p_{01}+p_{10}-p_{12}-p_{21})^2 +4p_{10}p_{12}}~.
\end{equation}
From this one obtains by means of equation \eref{eq:gantm_formel}
\begin{eqnarray}
\fl P_{11}(\tau) = \frac{p_{01}p_{21}}{\mu_1\mu_2} -
\frac{\e^{\mu_1\tau}}{\mu_1(\mu_1-\mu_2)}\Bigl(p_{10}(p_{21}+\mu_1) +
p_{12}(p_{01}+\mu_1)\Bigr)\nonumber\\
+ \frac{\e^{\mu_2\tau}}{\mu_2(\mu_1-\mu_2)}\Bigl(p_{10}(p_{21}+\mu_2) 
+ p_{12}(p_{01}+\mu_2)\Bigr)\\
\fl P_{12}(\tau) =  \frac{p_{01}p_{12}}{\mu_1\mu_2} +
\frac{p_{12}\e^{\mu_1\tau}}{\mu_1(\mu_1-\mu_2)}(p_{01}+\mu_1) -
\frac{p_{12}\e^{\mu_2\tau}}{\mu_2(\mu_1-\mu_2)}(p_{01}+\mu_2)\\ 
\fl P_{21}(\tau) = \frac{p_{01}p_{21}}{\mu_1\mu_2} +
\frac{p_{21}\e^{\mu_1\tau}}{\mu_1(\mu_1-\mu_2)}(p_{01}+\mu_1) -
\frac{p_{21}\e^{\mu_2\tau}}{\mu_2(\mu_1-\mu_2)}(p_{01}+\mu_2)\\ 
\fl P_{22}(\tau) = \frac{p_{01}p_{12}}{\mu_1\mu_2} +
\frac{p_{21}\e^{\mu_1\tau}}{\mu_1(\mu_1-\mu_2)}(p_{12}+p_{21}+\mu_2) -
\frac{p_{21}\e^{\mu_2\tau}}{\mu_2(\mu_1-\mu_2)}(p_{12}+p_{21}+\mu_1) 
\end{eqnarray}
and, for $\tau \rightarrow \infty$, by means of equation \eref{eq:P_i}
\begin{eqnarray}
P_0 &=& \frac{p_{10}p_{21}}{p_{10}p_{21} + p_{01}p_{12}+p_{01}p_{21}} \\
P_1 &=& \frac{p_{01}p_{21}}{p_{10}p_{21} + p_{01}p_{12}+p_{01}p_{21}} \\
P_2 &=& \frac{p_{01}p_{12}}{p_{10}p_{21} + p_{01}p_{12}+p_{01}p_{21}}~.
\end{eqnarray}

For two dipole-dipole interacting $V$ systems with  metastable state
(\fref{fig:V-schema}) the $p_{ij}$ have been calculated in
\cite{Addicks-EPJ-2001}. For this $V$ system  
the dipole-dipole coupling constant $C_2$
can be replaced by 0 since $A_2$ is very small. For simplification we
have also put $A_2=0$ which does not change the overall results. 
For zero detuning the transition rates $p_{ij}$ between the three
periods are then given in \cite{Addicks-EPJ-2001} as
\begin{eqnarray}
p_{01} &=&\frac{2A_3\Omega_2^2}{\Omega_3^2}\\
p_{10} &=&\frac{A_3^3\Omega_2^2}{(A_3^2+2\Omega_3^2)\Omega_3^2}\\
p_{12} &=&\Omega_2^2\left[\frac{A_3}{\Omega_3^2} + \mathrm{Re}\,
    C_3\frac{2A_3^2}{(A_3^2 + 2\Omega_3^2)\Omega_3^2}\right]\\
p_{21} &=&\Omega_2^2\left[\frac{2A_3^3}{(A_3^2 +
      2\Omega_3^2)\Omega_3^2} + \mathrm{Re}\,C_3 \frac{4A_3^4
      (A_3^2+4\Omega_3^2)}{(A_3^2 + 2\Omega_3^2)^3
      \Omega_3^2}\right]\nonumber\\ &&
\end{eqnarray}
to second order in $\Omega_2$ and first order in
$C_3$, with the remaining $p_{ij}$ being zero. The calculation of the
expressions to all orders in $C_3$ and for nonzero detuning is also given in
 \cite{Addicks-EPJ-2001}. For the period durations one has $T_0 =
1/p_{01}$, $T_1 = 1/(p_{10} + p_{12})$ and $T_2 = 1/p_{21}$. 
It is noteworthy that $P_i$ becomes independent
of $\Omega_2$, up to second order and to all orders of $C_3$.

\end{appendix}


\end{document}